\documentclass[10pt,letterpaper]{article}
\usepackage{opex3}

\usepackage{amsmath}
\usepackage{amssymb}
\usepackage{lastpage} 
\usepackage{multirow}
\usepackage{graphicx}
\usepackage{sidecap}

\begin{document}

\title{Parametric seeding of a microresonator optical frequency comb}

\author{Scott B. Papp, Pascal Del'Haye, and Scott A. Diddams}
\address{National Institute of Standards and Technology, Boulder, Colorado 80305, USA}
\email{scott.papp@nist.gov}

\begin{abstract}
We have investigated parametric seeding of a microresonator frequency comb (microcomb) by way of a pump laser with two electro-optic-modulation sidebands. We show that the pump-sideband spacing is precisely replicated throughout the microcomb's optical spectrum, and we demonstrate a record absolute line-spacing stability for microcombs of $1.6\times10^{-13}$ at 1 s.  The spectrum of a parametric comb is complex, and often non-equidistant subcombs are observed.  Our results demonstrate that parametric seeding can not only control the subcombs, but can lead to the generation of a strictly equidistant microcomb spectrum.   
\end{abstract}

%\bibliographystyle{/Volumes/work1/bib_files/apsrev4-1}
%\bibliography{/Volumes/work1/bib_files/sp_TF}

\bibliographystyle{../../../../bib_files/osajnl}
\bibliography{../../../../bib_files/sp_TF,../../../../bib_files/sp_QOpt}

The optical spectrum of a microresonator frequency comb (microcomb) is generated via parametric nonlinear optics, which can initiate with only milliwatts of CW pump laser power due to the high $Q$ and small mode volume of microresonators.  The simplicity of the generation mechanism, in combination with uniquely large line spacings in the 10's to 100's of GHz and the potential for photonic integration, makes microcombs a potential alternative for applications that currently rely on table-top combs. For example, applications such as direct spectroscopy \cite{Gerginov2005}, real-time trace detection \cite{Thorpe2006}, and astronomical spectrograph calibration \cite{Steinmetz2008,Ycas2012} call for robust, portable comb operation. The development of portable microcomb technology could also expand the reach of a comb's natural function, as an optical clockwork that leverages the remarkable precision of optical frequency measurements.

Microcomb generation has been studied with a range of microresonators from bulk devices such as microtoroids \cite{DelHaye2007}, crystalline resonators \cite{Savchenkov2008,Grudinin2009}, and microrods \cite{Papp2011,Papp2012,DelHaye2012} to integration-capable platforms such as silica \cite{Li2012a} and silicon nitride \cite{Levy2010}. A common observation in microcomb experiments has been the lack of a deterministic modelocking mechanism, which enables the type of stable operation, equidistance of all comb lines, and ultrashort optical waveforms known from femtosecond modelocked lasers. Still, a number of interesting microcomb operating regimes have been reported including low-noise \cite{Savchenkov2008a,DelHaye2008,Papp2012,DelHaye2012,Li2012a}, phaselocked \cite{Ferdous2011,Papp2011}, and stabilized operation \cite{DelHaye2008,Papp2012,DelHaye2012}, and the generation of soliton waveforms \cite{Herr2012a,Saha2013}.  Theoretical models have been proposed that explain many microcomb behaviors \cite{Herr2012,Matsko2005,Chembo2010}. But at present, a comprehensive picture of microcomb generation, and specifically a singular modelocking mechanism, have not been identified.

\begin{figure}[th]\centering
\includegraphics[width=0.75\textwidth]{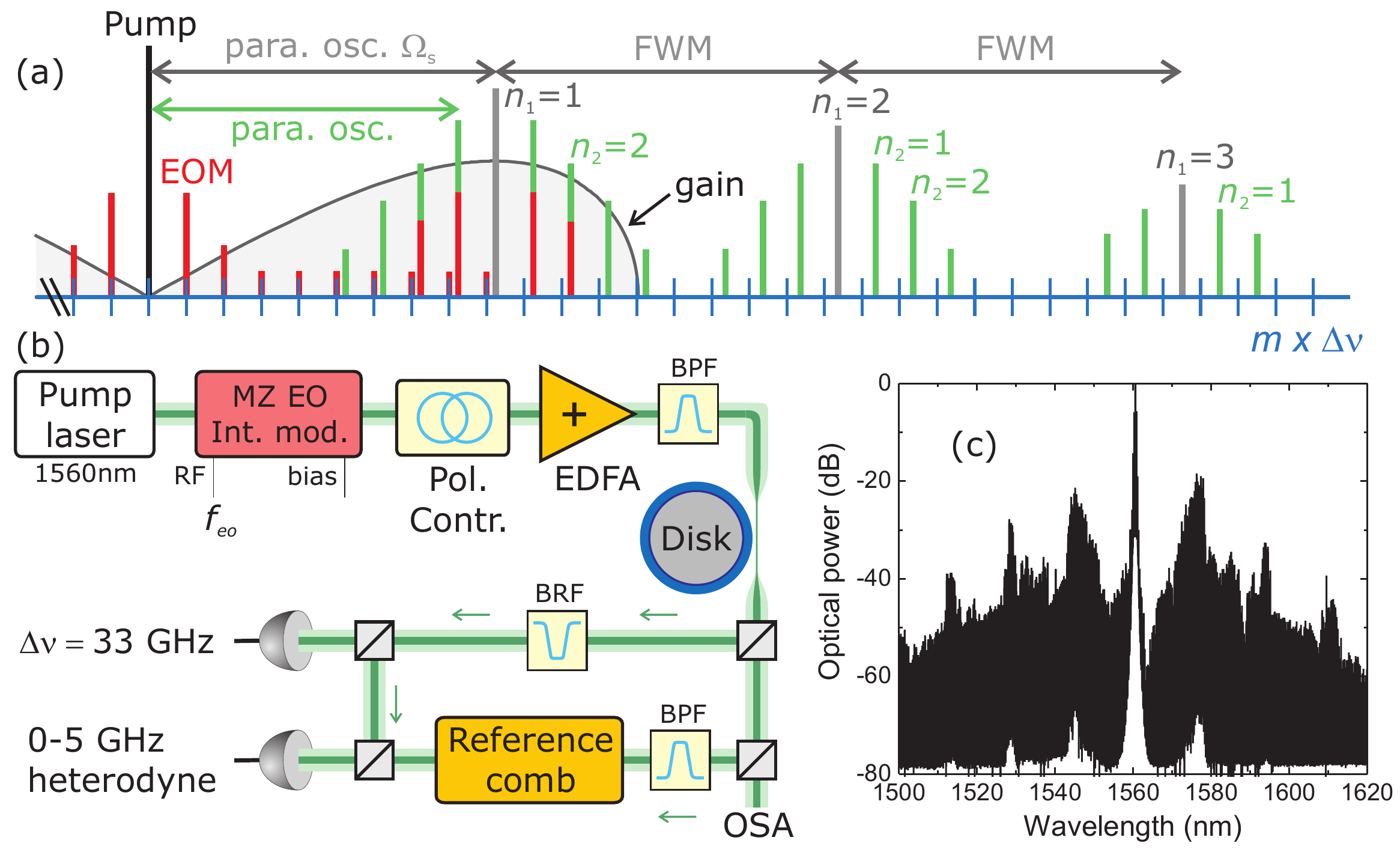}
\caption{Implementation of parametric seeding. (a)  Model for comb generation based on parametric oscillation and FWM.  A microcomb with a line spacing close to the resonator FSR is created via a primary parametric process ($n_1$) with spacing $\Omega_s\approx m_i$FSR in combination with a secondary process ($n_2$) active in adjacent resonator modes.  All other lines from these initial ones.  Parametric seeding can influence comb generation not only by initiating the secondary process, but via injection seeding of the primary one. (b)  The setup for our experiments consists of a laser, which is intensity modulated in a Mach-Zehnder EOM device, amplified, and coupled into a silica disk resonator \cite{Lee2012} via tapered fiber \cite{Cai2000}.  To analyze microcomb spectra, we use photodetection to record the line spacing, and we make line-by-line optical frequency measurements with respect to a reference comb.  BPF: bandpass filter, BRF: bandreject filter, OSA: optical spectrum analyzer. (c)  Optical spectrum for a microcomb with parametric seeding.
\vspace{-12pt}\label{fig1}}\end{figure}

This paper introduces a coherent control technique, \textit{parametric seeding}, for microcombs by way of pump-laser electro-optic (EO) intensity modulation. When the seeding frequency ($f_{eo}$) is tuned close to the microresonator free-spectral range (FSR), the intensity modulation strongly influences the parametric nonlinear processes responsible for comb generation. We observe that the $\sim33$ GHz line spacing ($\Delta\nu$) of the resulting microcomb spectrum is precisely locked to $f_{eo}$, and that $\Delta\nu$ can be varied over a range of $\sim10$ MHz, and we measure a record absolute $\Delta\nu$ stability of $1.6\times10^{-13}$ at 1 s, which is consistent with state-of-the-art microwave oscillators.  Furthermore, our parametric seeding technique enables systematic investigations of the microcomb generation process, specifically their non-equidistant ``sub-comb'' behavior first described in Ref. \cite{Herr2012}.  Finally, we demonstrate the emergence of a continuously equidistant microcomb spectrum tied to $f_{eo}$ that results from amplification of the seeding input signal.

Figure \ref{fig1}a presents a simple model for parametric comb generation in microresonators, which is motivated by Ref. \cite{Herr2012}.  A CW pump laser ($\omega_p$) excites one microresonator mode.  At the threshold power for parametric oscillation, signal and idler waves are generated at a frequency offset from the pump laser of $\Omega_s \sim m_i \times$FSR, where $m_i$ is an integer that counts by how many FSR the initial signal/idler pair are spaced from the pump.  The frequency $\Omega_s$ is determined by the balance of phase mismatch from dispersion, nonlinear effects, and the resonator mode structure; Fig. \ref{fig1}a shows the basic shape of single-pump parametric gain for the material dispersion of fused silica.  Without increased pump power, four-wave mixing of pump, signal, and idler creates fields at $\omega_p + n_1 \, \Omega_s$ (gray lines), where the integer $n_1$ indicates the primary spacing.  When pump power is increased, secondary parametric oscillation (often in resonator modes adjacent to the first signal/idler) and four-wave mixing (FWM) occurs, leading to a comb spectrum (green lines) with line spacing close to the resonator FSR.  Within the framework of this model the microcomb's spectrum is characterized by subcombs \cite{Herr2012} that are written as
\begin{equation} 
\nu_m=n_1 \, \Omega_s /2\pi + n_2 \, \Delta\nu,
\label{sub}\end{equation}
where $\nu_m$ is the frequency difference of $\omega_p/2\pi$ and the comb line, $m=n_1 m_i+n_2$ counts the mode number of the line relative to the pump, and $n_1$ ($n_2$) labels the order of the primary (secondary) parametric process.  Here the spacing between most lines is $\Delta\nu$, but the comb is not equidistant since an integer ratio for $\Omega_s/2\pi\Delta\nu$ is not required.  This peculiar spectrum can be compared to a completely equidistant one with line number $m$ separated from the pump laser by precisely $\nu_m = m \, \Delta\nu$ (blue axis markers).

\begin{figure}[t]\centering
\includegraphics[width=0.75\textwidth]{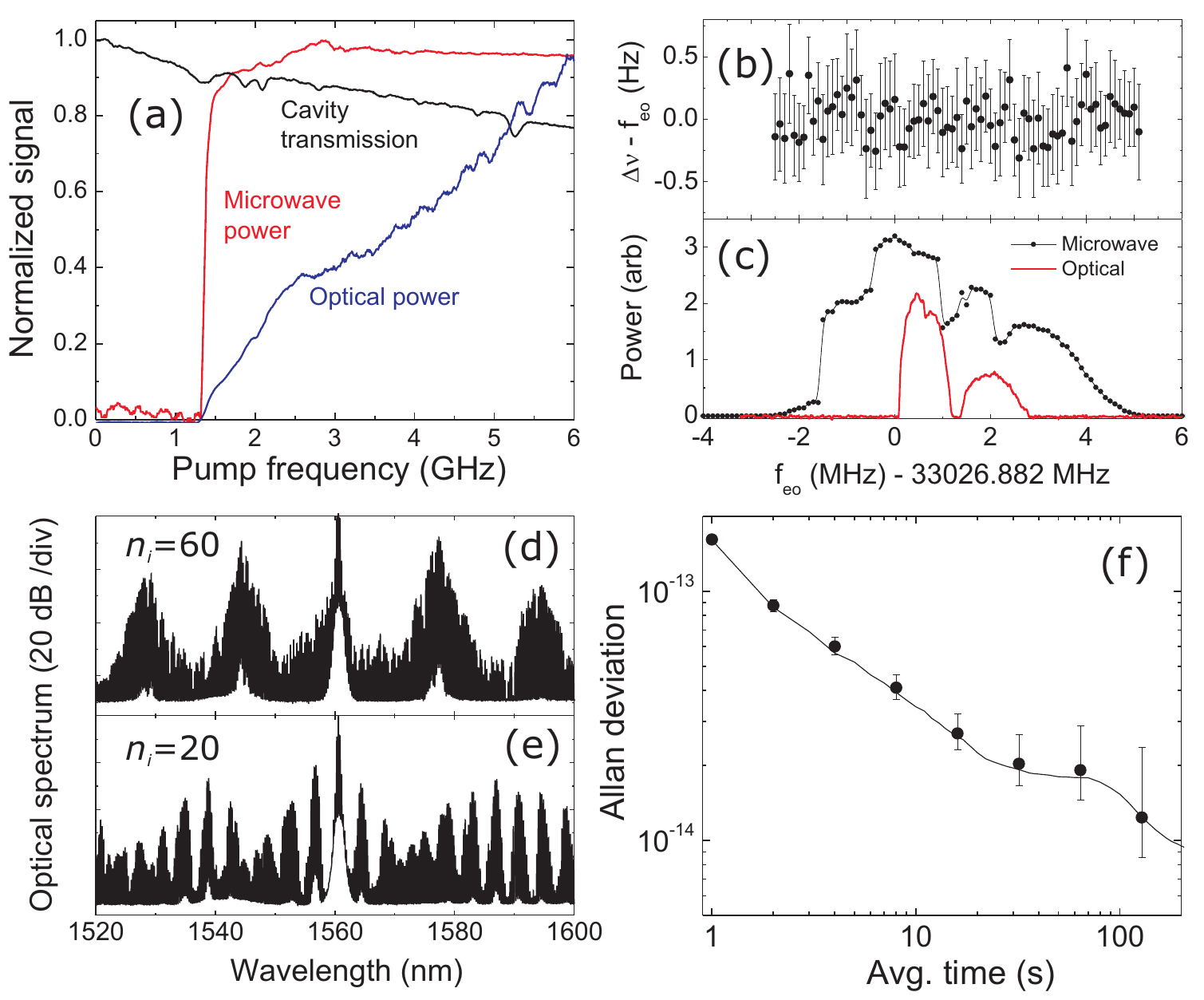}
\caption{Control of the microcomb line spacing via parametric seeding.  (a)  Disk resonator transmission (black), optical power (blue), and microwave power (red) as the pump laser is tuned onto resonance of the microcavity.  The optical and microwave outputs of the comb are measured after filtering out the pump and sidebands.   (b)  Near-zero difference between the comb's line spacing ($\Delta\nu$) and seeding frequency ($f_{eo}$).  (c)  Optical (red) and microwave (points) power as $f_{eo}$ is scanned.  The resonator FSR is 33.02932(9) GHz.  (d, e) Microcomb optical spectra for different settings of the seeding frequency; $f_{eo}=$33.026861 in (d) and $f_{eo}=$33.030 in (e).  (f)  Absolute line spacing Allan deviation as a function of measurement time.
\vspace{-12pt}\label{fig2}}\end{figure}

We expect parametric seeding with intensity modulation sidebands will impact microcomb generation in three ways.  First, a few lines near the pump laser are created by way of FWM of the pump and sidebands. These lines are strictly equidistant from the pump laser due to energy conservation of FWM; such a bichromatic pump was explored in Ref. \cite{Strekalov2009}.  However, parametric gain near the pump is low, which limits the number and output power of these new comb lines.  Second, pump intensity modulation creates sidebands of the parametric oscillation signal/idler waves at $\Omega_s$.  Since parametric gain is largest near the first signal/idler pair, the modulation-induced sidebands near $\Omega_s$ can rapidly grow, and further FWM processes also create new comb components, which importantly are all spaced by $f_{eo}$.  Third, the equidistant series of lines emanating from pump-sideband mixing can influence parametric oscillation at $\Omega_s$, and at best can completely injection-lock the primary process.  Our paper explores these three impacts of parametric seeding, and we identify some common behaviors in microcomb generation, specifically that a single microcomb mode is often a composite of multiple subcombs.

\vspace{-6pt}\section{Control of microcomb line spacing via parametric seeding}\vspace{-6pt}
 
A schematic of our system is shown in Fig. \ref{fig1}b.  To derive the parametric seeding signal, a CW laser's intensity is modulated at frequency $f_{eo}$ \footnote{The microwave synthesizer that generates $f_{eo}$ is referenced to a hydrogen-maser oscillator with $1.6\times10^{-13}$ 1 sec fractional frequency stability.} with a standard Mach-Zehnder electro-optic modulator.  The relative power of the carrier and seeding sidebands is controlled by a bias voltage, and is typically set to $>10\%$ of the carrier.  The second-order sidebands have 40 dB less power than the first-order ones.  After amplification to a maximum of 140 mW, the light is coupled into a disk resonator using a tapered fiber.  The 2 mm diameter, 8 $\mu$m thick silica disk resonator is fabricated on a silicon chip using the procedure described in Ref. \cite{Lee2012}.  The particular optical mode we use offers among the widest comb span ($\sim400$ nm) observed with this device.  By tuning the pump-laser frequency toward cavity resonance, we generate the gap-free optical spectrum shown in Fig. \ref{fig1}b that covers significantly more than 120 nm.  We analyze the microcomb spectrum with two techniques:  Direct photodetection to characterize the microwave line spacing, and by use of a reference comb (described below) to understand the frequency spacing between the pump and microcomb lines.  For all of the data in this paper except optical spectra, the central 2 nm of the comb is blocked to reveal information about only the parametrically generated comb.

Parametric seeding can significantly reduce the dramatic changes in optical and microwave outputs that are typically observed in single-pump microcombs \cite{Papp2011}.  Fig. \ref{fig2}a shows how the total cavity transmission (black line), integrated optical power (blue line), and integrated microwave power near 33 GHz (red line) of the comb evolve as the pump laser frequency is scanned.  Importantly at the $\sim$10 mW parametric threshold, both the optical and microwave power switch on.  In contrast, without $f_{eo}$ the lines spacing at threshold would be 60$\times$FSR and undetectable electronically.  In addition, the optical and microwave power outputs are stabilized by the parametric seeding, and they grow smoothly as the pump frequency tunes toward the microcavity resonance.

By tuning the seeding frequency $f_{eo}$, different aspects of parametric comb generation process can be explored.  Specifically, we discovered that distinct microcomb operating regimes exist, but that the line spacing is always precisely $f_{eo}$.  The red line in Fig. \ref{fig2}c shows the comb's optical power (central 2 nm about pump removed) as a function of $f_{eo}$.  Here the pump power is maintained below parametric threshold, and the comb's output is zero unless $f_{eo}$ is tuned onto resonance of modes adjacent to the pump.  Qualitatively different combs are generated for $f_{eo}$ settings only a few MHz apart: At $f_{eo}\lesssim 33.028$ GHz, the primary parametric process occurs at $m_i\approx60$ modes from the pump, while at $f_{eo}\gtrsim 33.028$ GHz, the comb is characterized by the smaller primary spacing $m_i\approx20$; and at the intermediate setting of $f_{eo}\approx 33.0282$ GHz no comb is generated.  For reference the resonator's FSR is 33.02932(9) GHz, which is measured at low power and calibrated with the modulation sidebands \cite{Li2012a}.  Parametric phase matching depends on numerous parameters, and is apparently influenced by the line spacing of the emergent comb.

When we increase intracavity power to above parametric threshold, the clean distinction between the two regimes is lost.  The black points in Fig. \ref{fig2}c show the complicated dependence of microwave output power on $f_{eo}$.  At the high and low $f_{eo}$ frequency tails of microwave power, we observe combs with either $m_i\approx60$ or $m_i\approx20$; the optical spectra of these cases (Fig. \ref{fig2}d and e) clearly show the different primary spacings.  But in between these extremes is a transition region in which both $m_i$ type combs are simultaneously possible.  Still, as shown in Fig. \ref{fig2}b, for every $f_{eo}$ data point the $\Delta\nu$ we measure (via frequency counting) is locked to $f_{eo}$ within $\pm0.5$ Hz.  Hence, no line-spacing stabilization post-generation is required.  For a single $f_{eo}$ setting, we measured the absolute Allan deviation of $\Delta\nu$ by comparing it to an independent hydrogen maser reference; see Fig. \ref{fig2}f.  The $1.6\times10^{-13}$ stability we achieve exceeds by 10x previous results obtained with microcombs \cite{Papp2012}.

\vspace{-6pt}\section{Optical frequency spectra of parametric subcombs}\label{subsec}\vspace{-6pt}

\begin{figure}\centering
\includegraphics[width=0.5\textwidth]{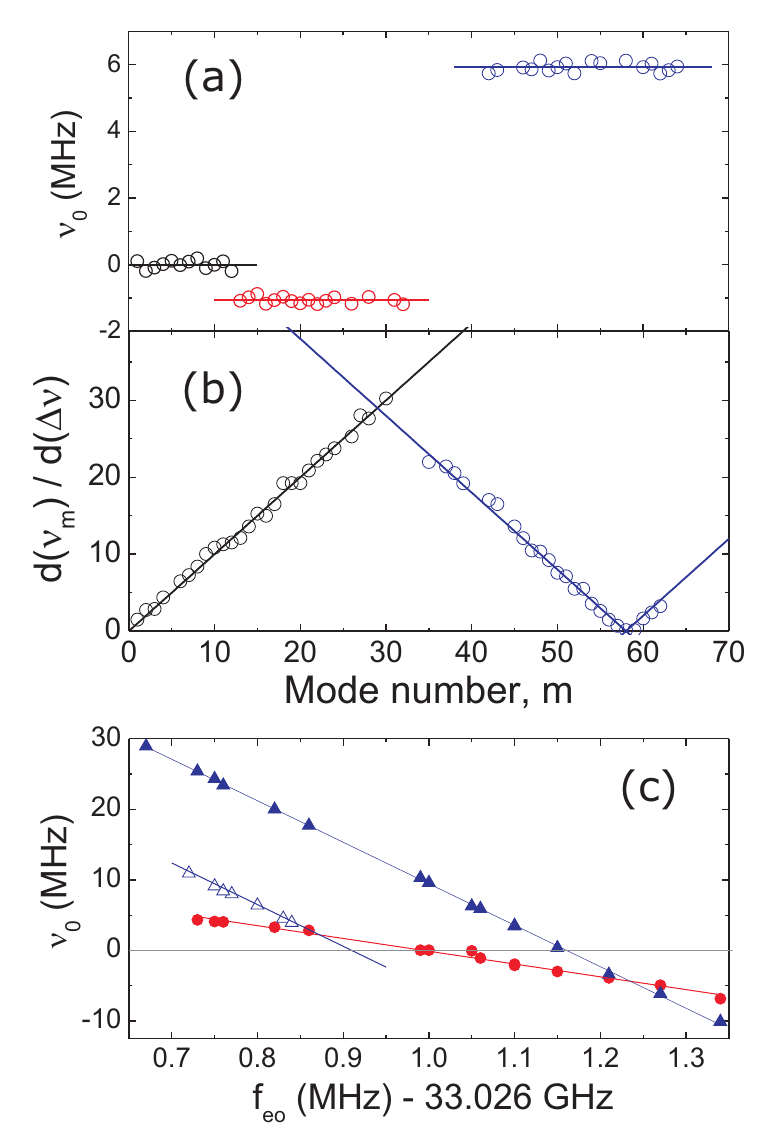}
\caption{Subcomb behavior in microcombs.  (a)  Line frequency diagram ($\nu_0$ vs. $m$) for a spectrum with three independent subcombs.  The black points represent FWM of the pump and seeding sidebands, and the red (blue) points represent the $m_i\approx20$ ($m_i\approx60$) subcomb.
(b)  Comb line tuning rate verses $m$.  Here $f_{eo}$ is set so only the $m_i\approx60$ subcomb is present.  (c)  Tuning of the independent subcomb offset frequencies with $f_{eo}$.  For the open triangles, the pump laser frequency was adjusted to shift the $\nu_0$ of all the points.
\vspace{-12pt}\label{fig3}}\end{figure}

Parametric seeding via intensity modulation has an impact not only on the microcomb line spacing, but also on the subcomb behavior described by Eqn. \ref{sub}.  Here we present direct tests of how subcomb lines depend on $f_{eo}$.  To characterize the optical frequency spectrum of our microcomb, we derive a reference comb from the pump laser; see the schematic in Fig. \ref{fig1}a.  After exiting the disk microresonator and tapered fiber, the pump laser is separated from the rest of the microcomb spectrum with an optical filter.  This pump light generates a comb covering 1540 nm to 1580 nm with 10 GHz spacing using a Fabry-Perot EO modulator \cite{Kourogi1993,Xiao2009}.  By photodetection of the interference between the microcomb and reference comb, we obtain a series of RF signals between zero and 5 GHz.  Each RF signal is associated with a specific optical heterodyne beatnote between microcomb and reference lines.  Using the two combs' precisely known line spacings, we determine the absolute difference between the pump and microcomb line ($\nu_m$), and the offset of the microcomb line from equidistance:
\begin{equation}\label{nu0}
\nu_0 = \nu_m-m \, \Delta\nu=n_1 \left(\Omega_s/2\pi -\Delta\nu \, m_i \right).
\end{equation}
Figure \ref{fig3}a shows $\nu_0$ versus microcomb mode number at $f_{eo}=33.02706$ GHz.  This microcomb line frequency diagram indicates the presence of three combs all with the same line spacing. \footnote{If the combs' spacing were not the same, then $\nu_0$ would vary with $m$.}  One comb (black points) originates directly from FWM of the pump laser and modulation sidebands, hence it has $\Omega_s=0$ and is strictly equidistant with offset -0.6 (120) kHz. But it is only composed of 13 lines near the pump.  The other two subcombs have offsets of -1.06 MHz and 5.93 MHz from equidistance, and they originate from independent parametric oscillation processes with $m_i\approx20$ (red points) and $m_i\approx60$ (blue points), respectively.  These data show that for this setting of $f_{eo}$, the microcomb operates in the transition between the two $m_i$ regimes, and that multiple independent parametric processes are active.

We analyze the subcombs' optical frequency spectrum by varying their line spacing with parametric seeding.  First, we focus on settings of $f_{eo}\lesssim 33.0269$ GHz for which the $m_i\approx60$ component is dominant.  To measure the tuning rate $|\frac{d(\nu_m)}{d(\Delta\nu)}|=|n_2|$, we vary $f_{eo}$ over a total range of 120 kHz and observe all the comb line frequencies for $m=\{0,70\}$; Fig. \ref{fig3}b shows the tuning rate versus mode number.  The blue points covering $m=\{35,65\}$ have a tuning rate characteristic of the $n_1=1$, $m_i\approx60$ subcomb, hence they have the expected $n_2=m-m_i$ dependence given by the blue lines.  This observation is strong evidence in support of the subcomb model, and future applications of microcomb technology could take advantage of the reduced sensitivity to $\Delta\nu$ of some lines.  Conversely, the black points at small $m$ result from FWM of the modulation sidebands and pump.  The linear increase in tuning rate with $m$ further confirms that the spectrum of this comb has $\Omega_s=0$ and its $n_2$ equals $m$.  

Second, we extract $\nu_0$ (Eqn. \ref{nu0}) from line-frequency diagrams taken over a wide range of $f_{eo}$.  Here both the $m_i\approx20$ and $m_i\approx60$ subcombs are excited.  The data showing offset from equidistance (Fig. \ref{fig3}c) for both the subcombs varies linearly with $f_{eo}$ with a slope of $-m_i$ and they are continuous through zero, as expected from the subcomb model.  Moreover, the $f_{eo}$ setting to realize $\nu_0 \simeq 0$ is different for the two subcombs.  Here the ability to systematically vary $\nu_0$ is primarily due to control of $\Delta\nu$, as opposed to a more fundamental ability to influence the parametric phase-matching condition, which determines $\Omega_s$.  Hence, when we change the intracavity power via the pump-laser frequency, which is expected to modify $\Omega_s$, we observe a shift in the $\nu_0=0$ intercept; see the open triangles in Fig. \ref{fig3}c.  However the underlying dependence of $\nu_0$ on $f_{eo}$ is unchanged.

\vspace{-6pt}\section{Emergence of a continuously equidistant microcomb}\vspace{-6pt}

\begin{figure}[t]\centering
\includegraphics[width=0.5\textwidth]{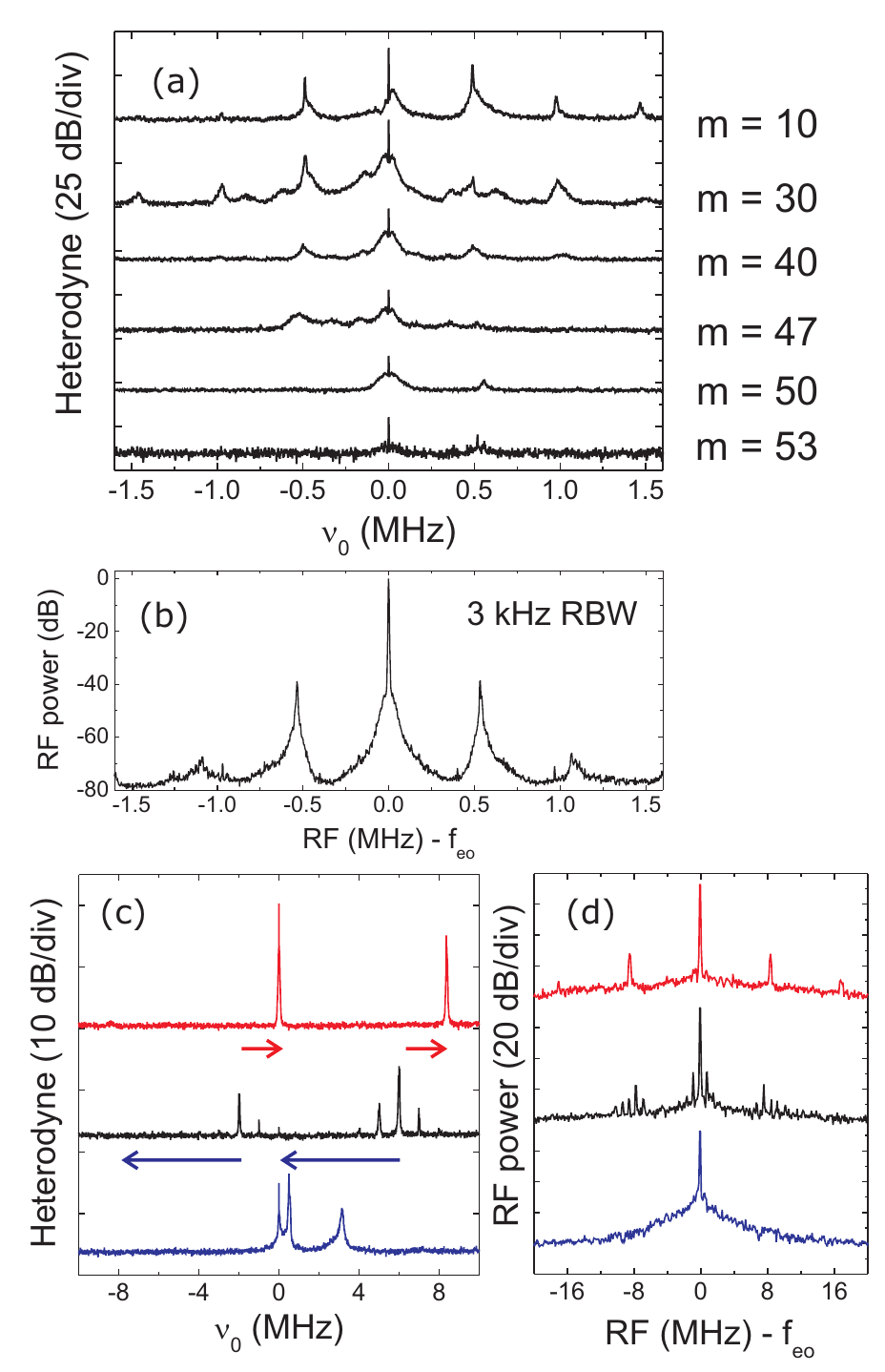}
\caption{Onset of a continuously equidistant microcomb.  (a) Optical heterodyne of various microcomb $m$-lines and the reference comb.  The narrow peak in each trace at $\nu_0$ indicates the presence of an equidistant microcomb line, while the other peaks belong to the $m_i\approx60$ subcomb.  (b)  Microcomb line spacing spectrum for data in (a).  The equidistant and subcomb lines have the same spacing, but the subcomb offset appears here at $\pm0.5$ Hz.  (c) Black trace is the optical heterodyne signal for microcomb line $m=40$ showing the $n_1=2$, $m_i\approx20$ subcomb at -2 MHz, and the $n_1=1$, $m_i\approx60$ subcomb at 5 MHz.  Apparently the microcomb's spectrum can consist of lines separated by more than the resonator's linewidth.  Tuning the pump laser frequency shifts the subcomb lines' $\nu_0$.  The upper (red) and lower (blue) traces show injection locking of the two subcombs, which is obtained by tuning their $\nu_0$ toward zero.  (d)  Microcomb output of the comb for the traces in (c). 
\vspace{-12pt}\label{fig4}}\end{figure}

By tuning the subcomb offset to specific values near zero, we observe the emergence of a continuously equidistant comb spectrum to at least microcomb mode $54$ ($\pm1.8$ THz from pump).  The offset-free comb lines, which are seeded by the intensity modulation sidebands, experience significant parametric amplification.  Such an equidistant comb appears even in the presence of the other $m_i\neq0$ subcombs that we described in Sec. \ref{subsec}, and multiple lines are observed in a single $m$-mode of the microcomb.  Figure \ref{fig4}a shows the heterodyne beat spectrum of the microcomb and reference comb for several $m$ values.  Here the frequency axis is precisely calibrated (at the Hz level on the underlying $\sim$THz pump-comb frequency difference) to represent the offset from equidistance.  Each trace is composed of several peaks, including a peak for the equidistant comb, a peak for the $m_i\approx60$ subcomb, and their nonlinear mixing products.  The peaks centered at $\nu_0=0$ have a narrow spectral width since pump laser frequency noise is common mode to both the equidistant microcomb and the reference comb.  Moreover the $\nu_0$ frequency of these peaks does not vary from zero when the pump frequency is tuned.  These factors help significantly in identification of the equidistant comb.  As expected, the $\nu_0$ frequency of lines shown here, which belong to the $m_i\approx60$ subcomb, do change with pump frequency.

The microwave line spacing spectrum we obtain via photodetection also provides information about the equidistant and offset microcomb components; see Fig. \ref{fig4}b.  Here the frequency axis covers the same range as in Fig. \ref{fig4}a to highlight their correspondence.  This signal is composed of a strong peak at $f_{eo}$ due to pairwise interference of adjacent lines among all the subcombs, and also of a series of peaks at multiples of 0.5 MHz from $f_{eo}$.  Previous microcomb experiments identified these additional peaks in the line-spacing spectrum as interference of overlapping subcomb lines with, for example, $n_1=1$ and $n_2=2$ \cite{Herr2012}.  However, in our case they also clearly indicate the presence of a continuously equidistant microcomb spectrum among separate, offset subcombs.  In fact, increasing the power in the additional peaks leads to a stronger equidistant comb.

\begin{figure}[t]\centering
\includegraphics[width=0.75\textwidth]{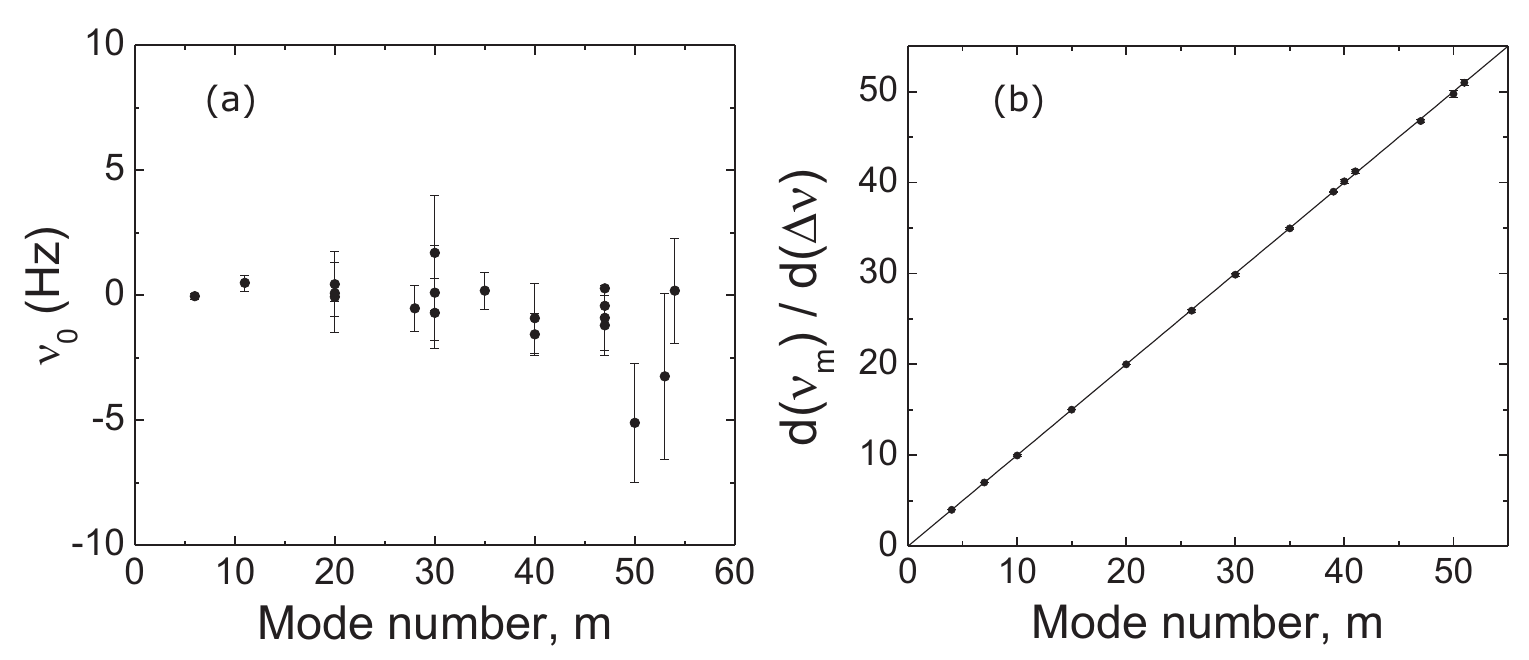}
\caption{Properties of the equidistant microcomb spectrum.  (a)  Measurements showing no offset from  equidistance at the 2 Hz level for select $m$ values up to 54.  Error bars are the standard deviation of the mean for each point.  (b)  Equidistant comb line tuning versus $m$ with slope of one.  
\vspace{-12pt}\label{fig5}}\end{figure}

We interpret the spectra in Fig. \ref{fig4}a and b as evidence for parametric-gain competition between different processes that generate the composite microcomb spectrum.  While the parametric gain spectrum is extremely broad, signal/idler fields can only be efficiently created within the resonance spectrum of the optical microcavity.  Naturally, to generate a completely equidistant comb the resonator modes must roughly line up with a fixed spacing; whereas for an offset subcomb, $\Omega_s$ of signal/idler fields changes with the resonator modes.  Hence, a subcomb line provides a coarse \textit{in-situ} marker of the resonator mode.  For a fixed setting of $f_{eo}$, we explore the competition between parametric processes by varying the pump laser frequency.  Fig. \ref{fig4}c shows heterodyne beatnote spectra of the $m=40$ microcomb with $\nu_0$ calibrated based on $f_{eo}$; the three traces are different settings of the pump laser frequency, and in Fig. \ref{fig4}d are the corresponding line-spacing signals.  The black trace (middle) demonstrates the complicated generation mechanism of parametric combs: The peaks at positive (negative) $\nu_0$ originate from a $m_i\approx60$ ($m_i\approx20$) subcomb, and their $\sim1$ MHz side peaks are associated with nonlinear mixing of their subcomb offset.  For the black trace, the pump frequency is set such that the equidistant $m=40$ line is between the two subcomb lines.  All the characteristics of this optical heterodyne signal are mirrored in the line spacing spectrum we measure, namely peaks at $f_{eo} \pm8$ MHz and $\pm1$ MHz corresponding to the subcomb offsets.  Conversely, when we use the pump laser to tune the $m_i\approx20$ line close to $\nu_0=0$ (upper, red trace), the $m=40$ equidistant component is strongly amplified, the $m_i\approx20$ subcomb component disappears, but the $m_i\approx60$ subcomb remains.  Tuning the pump frequency in the opposite direction (lower, blue trace) results in amplification of the equidistant line, reduction of the $m_i\approx60$ subcomb, and complete disappearance of the $m_i\approx20$ one.  Again, the microwave line spacing spectra we obtain mirror the behavior of corresponding red and blue optical heterodyne traces.

Generating an equidistant spectrum is important for many future applications of microcombs, particularly ones that leverage frequency division from the optical to microwave domains \cite{Fortier2011}.  To characterize its equidistance, we perform high-resolution measurements of the microcomb optical frequency spectrum using the reference comb.  Here we use a microwave spectrum analyzer to repeatedly measure the microcomb-reference beat frequencies.  Figure \ref{fig5}a shows measurements of $\nu_0$ for up to mode $54$ from the pump laser.  The offset we observe for all the studied $m$ values is within the $\sim2$ Hz uncertainty of our measurements.  Specifically, for $m=54$ the fractional uncertainty in our offset measurement is $1\times10^{-12}$.  By varying $f_{eo}$ and monitoring the frequency of each comb line, we characterize the tuning behavior of the equidistant comb; see Fig. \ref{fig5}b.  In contrast to the subcomb measurements presented in Sec. \ref{subsec}, the equidistant comb tunes precisely with the expected dependence on $m$.  These data demonstrate that, even for the weak -60 dBm optical line powers of our reference comb, the equidistant microcomb we create is capable of precision optical frequency metrology across a significant fraction of its span.

\vspace{-6pt}\section{Conclusion}\vspace{-6pt}

In this paper, we have introduced a new parametric seeding technique for microcombs that provides for deterministic generation of the comb spectrum with known spacing.  Our measurements of the absolute line-spacing stability reach a record level commensurate with state-of-the-art electronic oscillators. Moreover, the data in this paper demonstrate that parametric seeding is an important tool for understanding the generation of microcombs.  Our simplified generation model (Fig. \ref{fig1}) is based on primary and secondary parametric oscillation processes, and we observe that at least three pairs of these processes can be active in creating a composite microcomb spectrum.  Importantly, parametric seeding can influence all aspects of microcomb generation.  We verified the dependence of subcomb lines on the microcomb's fundamental line spacing, and we demonstrated that subcomb offsets can be tuned through zero by changing the line spacing.  However, simply tuning the subcomb offset to zero with $\Delta\nu$ does not necessarily yield the type of equidistant comb that will be important for future applications.  On the other hand, even in the presence of multiple subcombs, we observe the emergence of a continuously equidistant spectrum that results from amplification of the parametric seeding signal, and we show that its offset from equidistance can be as low as $10^{-12}$.  Future experiments will focus on seeding with more than two sidebands, and on investigating $\sim$picosecond optical waveforms associated with the microcomb spectra presented here.

We are grateful to Hansuek Lee, Jiang Li, and Kerry Vahala for providing the silica disk resonator.  We thank Lora Nugent-Glandorf and Frank Quinlan for providing helpful comments on this manuscript.  This work is supported by the DARPA QuASAR program and NASA.  It is a contribution of the US government (NIST) and is not subject to copyright in the United States.

\end{document}